\documentclass[aps,physrev, twocolumn,superscriptaddress]{revtex4-2}

\usepackage{parskip}
\usepackage{graphicx}
\usepackage{color}
\usepackage[normalem]{ulem}

\usepackage{amsmath}
\usepackage{hyperref}
\usepackage{comment}

\begin{document}

\title{Non-equilibrium geometric forces steer spiral waves on folded surfaces}

\author{Varun Venkatesh}
\affiliation{Niels Bohr Institute, University of Copenhagen, 2100 Blegdamsvej 17, Copenhagen, Denmark}
\author{Farzan Vafa}
\affiliation{Niels Bohr Institute, University of Copenhagen, 2100 Blegdamsvej 17, Copenhagen, Denmark}
\affiliation{Department of Physics, Massachusetts Institute of Technology, Cambridge, MA, USA}
\author{Martin Cramer Pedersen}
\affiliation{Niels Bohr Institute, University of Copenhagen, 2100 Blegdamsvej 17, Copenhagen, Denmark}
\author{Amin Doostmohammadi}
\email[]{doostmohammadi@nbi.ku.dk}
\affiliation{Niels Bohr Institute, University of Copenhagen, 2100 Blegdamsvej 17, Copenhagen, Denmark}

\date{\today}

\begin{abstract}
Spiral waves are ubiquitous signatures of non equilibrium dynamics, appearing across chemical, biological, and active systems. Yet, in many living systems these waves unfold on curved and folded surfaces whose geometry has rarely been treated as a dynamical factor. Here we show that surface curvature fundamentally shapes spiral wave behavior and can contribute to the organization of neural activity in the brain. Via analytical theory and simulations of the complex Ginzburg Landau equation (CGLE) on curved surfaces, we demonstrate that curvature enters through the Laplace Beltrami operator as a spatial modulation of effective diffusion. Gradients of this effective diffusion generate a geometric force on spiral defects, and the complex nature of the CGLE produces a complex mobility that leads to non central and non reciprocal responses. Applied to realistic cortical surfaces of the human brain, the model predicts that the pattern of cortical folding stabilizes and localizes spiral waves, while progressive smoothing of the surface erases these non equilibrium structures. This reveals that brain geometry is not a passive scaffold but an active physical constraint that shapes neural dynamics. More broadly, the same geometric mechanism provides a universal route by which curvature and topology control pattern formation across oscillatory, chemical, and active matter systems.
\end{abstract}
\maketitle

\section{Introduction \label{intro}}
Spiral waves are among the most ubiquitous and visually striking examples of non equilibrium pattern formation in nature, emerging spontaneously in diverse systems where excitable or oscillatory media are driven far from equilibrium. They have been observed across a remarkable range of systems, from chemical reactions to biological tissues~\cite{aranson2002world}. In the Belousov-Zhabotinsky reaction they appear as rotating reaction fronts visible to the naked eye~\cite{zaikin1970concentration,keener1986spiral}, while in colonies of \textit{Dictyostelium} slime molds they coordinate the aggregation of millions of cells through concentric and spiral waves of cAMP signaling~\cite{siegert_spiral_1995, palsson_origin_1996}. In living tissues, spiral waves of electrical or biochemical activity are closely linked to both normal physiological function and pathological states. They underlie life threatening cardiac arrhythmias~\cite{davidenko_stationary_1992,hwang_complex-periodic_2005}, organize morphogenetic signaling during development~\cite{lechleiter_spiral_1991}, and shape the spatiotemporal activity of the cerebral cortex~\cite{huang_spiral_2004, liang_complexity_2023, xu2023interacting}. This remarkable universality suggests that spiral waves represent a fundamental organizing motif of non equilibrium media and provide a unifying language for describing dynamical order across physics, chemistry, and biology.

Central to this organizing role are the spiral centers, or topological defects, around which wave fronts rotate~\cite{ardavseva2022topological}. These point singularities act as organizing centers for the surrounding medium, often determining its global dynamics. In cardiac tissue, the trajectories of spirals decide whether an arrhythmia self terminates or degenerates into fibrillation~\cite{christoph2018electromechanical, kim_cardiac_2007}. In cortical tissue, recent work shows that spiral defects mark hubs of information flow and coincide with regions of strong coupling between large scale functional networks~\cite{xu2023interacting}. Despite the universal appearance of spiral waves in biological settings, theoretical descriptions overwhelmingly assume flat, homogeneous domains~\cite{neu1990vortices, aguareles2010motion,aguareles2020dynamics}. Real biological systems, however, are far from flat. The cortex is a folded surface with strongly varying curvature, the heart is a curved and anisotropic shell, and developing tissues continuously deform as they grow. How the intrinsic geometry and topology of these curved and often highly folded surfaces influence spiral wave dynamics therefore remains a fundamental and largely unanswered question at the interface of physics, mathematics, and biology.

 \begin{figure*}[t!]
    \centering
    \includegraphics[width=0.95\linewidth]{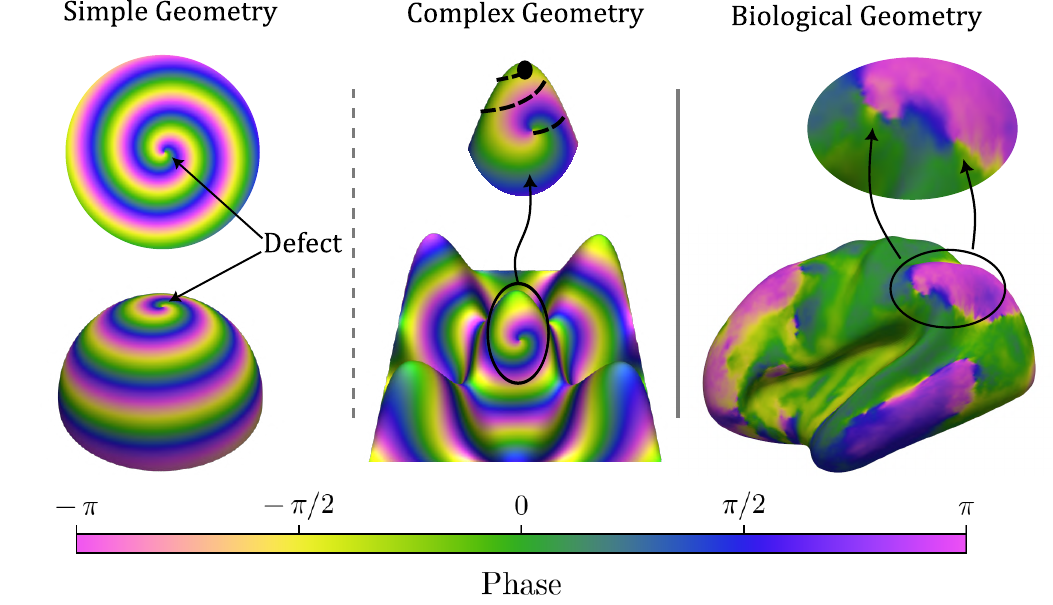}
    \caption{\textbf{Curvature determines spiral wave defect dynamics.} Simulations of spiral waves on simple surfaces, such as disk and spherical caps, exhibit equilibrium steady state defect configurations (left). On surfaces with complex curvature landscapes like a sinusoidal geometry, defects undergo non equilibrium dynamics (center). Experimental observations on the cortical surface reveal motile defects with trajectories spanning cortical regions (right). Experimental data obtained from Ref.~\cite{doi:10.1126/sciadv.abf2709}.}
    \label{fig:fig1}
\end{figure*}

The importance of this question has been amplified by a series of recent experimental discoveries. In the human brain, interacting spiral waves have been shown to be widespread modes of activity during rest and cognition, where they coordinate large scale patterns of neural excitation and flexibly route information between cortical networks~\cite{xu2023interacting}. During sleep, rotating waves and spiral like spindles organize global neocortical dynamics and predict the consolidation of memory traces over hours~\cite{muller_rotating_2016, xu_spatiotemporal_2025}. In mice, high resolution mesoscopic imaging and large scale electrophysiology have revealed brain wide spiral waves whose centers align with specific cortical regions and even with the local axonal architecture, which exhibits matching circular organization~\cite{ye_brain-wide_2023}. Together, these findings indicate that spiral waves are robust, large scale organizing modes that structure brain dynamics across species, behavioral states, and spatial scales. In parallel, new theoretical and empirical work on geometric eigenmodes has demonstrated that the geometry of the brain, including its folding pattern and curvature, can be a more fundamental determinant of macroscopic activity than the detailed pattern of long range connectivity~\cite{pang2023geometric}. Yet, analyses of brain spirals almost universally project cortical activity onto flat or heavily smoothed surfaces, effectively removing the very geometric features that may be essential for generating and sustaining these dynamical patterns. In this context, the brain serves as a stringent and biologically relevant test bed for a general physical principle: how geometry controls nonequilibrium dynamics.

Related advances in other pattern forming systems have demonstrated that curvature can actively drive dynamics rather than merely provide a passive substrate. In nematic liquid crystals, Gaussian curvature alone generates forces that propel topological defects into orbits, bias their creation and annihilation, and produce complex periodic motion on curved shells~\cite{turner2010vortices,vafa2022defectAbsorption,vafa2025defect,vafa2023active,vafa2024periodic}. In reaction diffusion media, curvature has been shown to trigger spontaneous motion of Turing and oscillatory patterns and to induce symmetry breaking in otherwise uniform systems~\cite{nishide_pattern_2022, stoop_curvature-induced_2015, nishide_oscillatory_2025}. These discoveries reveal that geometric fields such as curvature can act as control parameters for non equilibrium pattern formation. However, spiral waves differ fundamentally from orientational or flow based systems because they couple amplitude and phase dynamics and generate long range wavefront interactions that feed back onto the defect core. As a result, the way curvature interacts with spiral waves cannot be inferred from analogies to Turing or hydrodynamic systems~\cite{turner2010vortices}. The few existing theoretical studies of spiral waves on curved surfaces have been restricted to simple geometries such as spheres~\cite{dai_ginzburg--landau_2021}. Such constant curvature settings eliminate curvature gradients and therefore miss the geometric forces identified here, leaving open whether more realistic and spatially heterogeneous surfaces, such as the folded cortex, give rise to qualitatively new behavior.

In this work, we move beyond the conventional flat space description and systematically investigate how intrinsic surface geometry governs spiral wave dynamics. We employ the complex Ginzburg Landau equation (CGLE), the universal amplitude equation near oscillatory instabilities, as a minimal framework that isolates purely geometric effects independent of microscopic biological detail~\cite{aranson2002world}.

Analytical theory and direct numerical simulations reveal that gradients in a geometric potential~\cite{vitelli2004anomalous,vafa2022defectAbsorption}, sourced by curvature, act through the Laplace Beltrami operator as spatially varying effective diffusion. These diffusion gradients generate a geometric force on defects. The complex mobility inherited from the CGLE produces both radial and transverse responses, yielding non central and non reciprocal motion. In analogy with electrostatics, curvature plays the role of charge density, and the field that it generates drives the defect with a complex mobility term. Applying this framework to realistic cortical surfaces, we find that gradually smoothing the brain geometry systematically erases defect localization and reduces their lifetimes, demonstrating that the folding pattern of the cortex can stabilize and organize spiral activity. This establishes surface geometry as a physical control parameter for spatiotemporal brain dynamics and provides a mechanistic link between the morphology of the brain and its emergent dynamical states. Because the underlying mechanism originates from the universal amplitude equation of oscillatory media, the principles we identify extend far beyond neuroscience, offering a general route by which curvature and topology can control pattern formation in chemical, biological, and active matter systems.

\section{Model \label{methods}}
We consider spiral waves generated by the complex Ginzburg Landau equation (CGLE) on curved surfaces, as this equation represents the fundamental, universal form governing spiral wave dynamics irrespective to the underlying system:
\begin{equation}
\frac{\partial \psi}{\partial t} = \psi + (1 + i\alpha)\nabla^2 \psi - (1 + i\beta)|\psi|^2 \psi \label{eq:CGLE}
\end{equation}
where $\psi(\mathbf{x},t)$ is a complex order parameter field representing the local amplitude and phase of the oscillations, $\alpha$ and $\beta$ characterize linear and nonlinear dispersion. Our paper focuses on both analytical and numerical investigations of this PDE, which we simulate on discrete triangle meshes, where the complete details can be found in the SI.

Following Ref.~\cite{aguareles2010motion}, we define the twist parameter $q \equiv (\beta - \alpha)/(1 + \alpha\beta)$, where $q$ characterizes the imaginary contributions of the coefficients of our system. In particular, increasing $q$ enhances the rotational frequency and reduces the wavelength of spiral waves~\cite{hagan1982spiral}. In the rest of the study we fix $q = -\alpha = -0.5$ unless otherwise specified.

\section{Gaussian curvature singularity induces spiral motion}

To understand the influence of surface geometry on spiral wave defects, we first examined defect dynamics on simple surfaces: flat planes and spherical caps. On both geometries, spiral wave defects reach equilibrium (Fig.~\ref{fig:fig1}). The uniform curvature of a spherical cap provides no preferred direction to break the rotational symmetry of the defect. In contrast, on more general surfaces such as sinusoidal geometries, defects exhibit rich and complex non equilibrium dynamics. Defects now move in response to the curvature landscape, spiraling away from regions of positive curvature towards regions of negative curvature, while displaying orbital motion around saddle points. The decisive distinction is the presence of curvature gradients, which act through the metric as gradients in effective diffusion and thereby generate a geometric force on defects. These geometric principles have direct consequences for understanding spiral wave behavior in biological systems. Spiral waves have been observed on incredibly complex surfaces such as the brain (Fig.~\ref{fig:fig1}), where they move and display trajectories spanning entire hemispheres~\cite{xu_spatiotemporal_2025}. For analytical tractability, we begin by studying the dynamics on a cone, where the Gaussian curvature is zero everywhere except for a singularity at the apex, providing a natural base for analytical work.

\begin{figure*}
    \centering
    \includegraphics[width=0.99\linewidth]{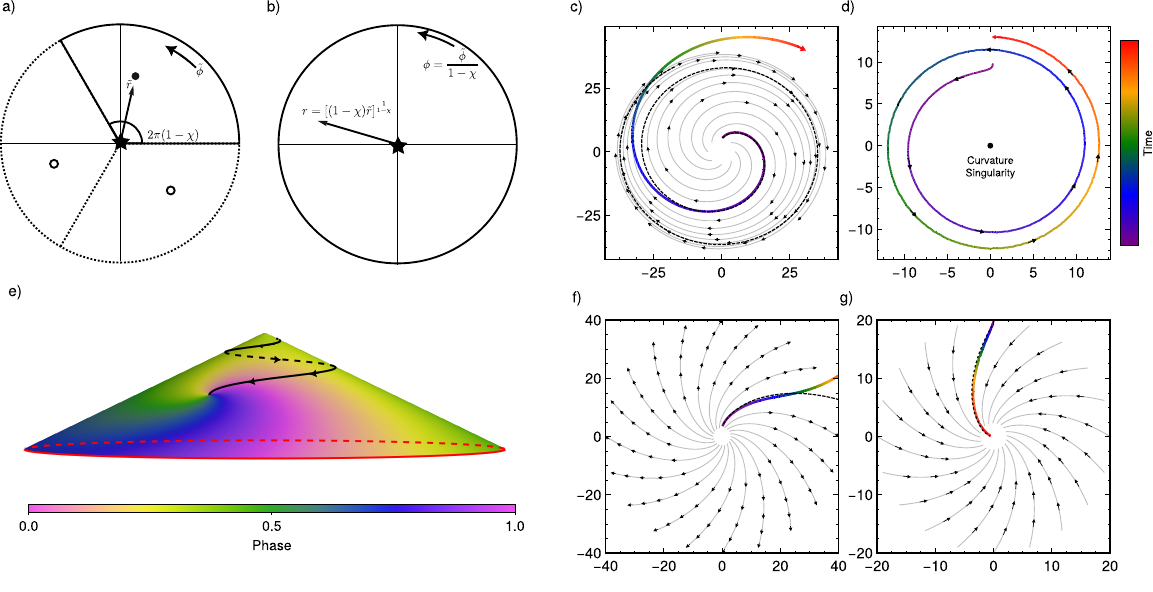}
    \caption{\textbf{Validation of near field theory on conical geometry.} (a) Unrolled flat wedge coordinates $(\tilde r, \tilde \theta)$, obtained by cutting and flattening the cone into the plane. The azimuthal angle spans $\tilde \phi \in [0, 2\pi(1-\chi))$, reflecting the angular deficit at the apex. The conical singularity is indicated by the star at the origin. A physical topological defect is shown as a black dot in the principle wedge, while the white circles in the dashed wedges represent the $n-1$ image defects; the example shown corresponds to $\chi = 2/3$ and $n = 3$. (b) Isothermal coordinates $(r,\theta)$ obtained by conformally mapping the cone to the plane such that the full azimuthal angle $\phi \in [0,2\pi)$ is restored. The relation between the two coordinate systems is indicated schematically, with the radial rescaling $r = [(1-\chi)\tilde r]^{1/(1-\chi)}$ and angular mapping $\phi = \tilde \phi/(1-\chi)$. With $\alpha = 0.5$ and $\chi = \pm 0.25$, analytical and simulated trajectories are compared for positive (c) but not negative (d) Gaussian curvature. Simulated defect trajectory for $\alpha=0.3$ on a cone corresponding to $\chi=0.1$ (e). At low parameters ($\alpha = 0.1$, $\chi = \pm 0.1$), analytical and simulated trajectories match remarkably well for both positive (f) and negative (g) Gaussian curvature.}
    \label{fig:cone_full}
\end{figure*}

For simplicity, we consider a cone where only a fraction $\frac{1}{n}$ of the full disk is present, i.e., a cone with deficit angle $2\pi\chi = 2\pi (1 - \frac{1}{n})$ (see Fig.~\ref{fig:cone_full}a, b for an introduction). We now take advantage of the method of images; namely, that we consider $n$ copies of the wedge, where on the $m$th copy the field is rotated by an angle $\frac{2\pi m}{n}$, for $m=1,\ldots,n-1$. In particular, this means that each defect has the same charge $\sigma$, and they are all situated on the vertices of a regular $n$-gon, where in complex coordinates the $m$th defect is at position $\tilde z_m = e^{2\pi i \frac{m}{n}} \tilde z$, where the defect is at $\tilde z = \tilde z_0$. 
We first express the defect velocity $\frac{d\tilde z}{dt}$ in flat wedge complex coordinates,
and then to compare to numerical simulations, we convert to isothermal coordinates, where $\frac{dr}{dt}$ and $\frac{d\theta}{dt}$ denote the radial and tangential components of the velocity of the defect. Fig.~\ref{fig:cone_full}a,b shows the transformation from polar physical to isothermal coordinates.

\subsection{Analytical treatment: near field}
Recent advances have provided analytical formulations for defect interactions in flat space for the complex Ginzburg-Landau equation~\cite{aguareles2010motion,aguareles2020dynamics}. Building on these works, we employ the near field dynamic equations for defects on flat geometry, which is valid for $q\ll 1$ and $0 < |q \log \epsilon| < \pi/2$ and adapt it for the cases when the geometry is not flat.

For $N$ spirals labeled by $\ell=1,\dots,N$, centered at positions $\mathbf{x}_\ell$, each carrying an integer winding (topological) charge $\sigma_\ell=\pm1$, the velocity of spiral $\ell$ due to all other spirals $j$, is given by:
\begin{align}
    \frac{d\mathbf{x}_\ell}{dt} =
    2 q \sigma_\ell\sum_{\substack{j=1 \\ j\neq \ell}}^N \bigl[&\sigma_j \cot (q |\log\epsilon(r_\ell)|)\nabla_\ell \log (x_{\ell j})\nonumber \\
    &+ \nabla_\ell^\perp \log(x_{\ell j})\bigr]
    \label{eqn:2defect}
\end{align}
where for notational convenience, $\nabla_\ell \equiv (\partial_{x_\ell}, \partial_{y_\ell})$,  $\nabla_\ell^\perp \equiv (-\partial_{y_\ell}, \partial_{x_\ell})$ and $x_{\ell j} \equiv |\mathbf{x}_\ell - \mathbf{x}_j|$.

For a cone, Eq.~\eqref{eqn:2defect} in complex coordinates $\tilde{z} = \tilde x + i\tilde y$ becomes
\begin{align}
    \frac{d\tilde z}{dt} &= 4q (\cot (q |\log\epsilon|) + i\sigma) \sum_{j=1}^{n-1} \left.\bar\partial\log |\tilde z - \tilde z_j|\right|_{\tilde z_j = \tilde z e^{2\pi i j/n}} \nonumber \\
    &= 2q(\cot (q |\log\epsilon|)  + i\sigma)\frac{\tilde z}{\tilde r^2}\sum_{j=1}^{n-1}\frac{1}{1 - e^{2\pi i j/n}},
    \label{eq:dztildedt_0}
\end{align}
where partial complex derivatives $\partial = \partial_z \equiv (\partial_x - i \partial_y)/2$ and $\bar\partial = \partial_{\bar z} \equiv (\partial_x + i \partial_y)/2$.
Using the below identity
\begin{equation}
\sum_{j=1}^{n-1}\frac{1}{1 - e^{2\pi i j/n}} = \frac{n-1}{2}
\end{equation}
Eq.~\eqref{eq:dztildedt_0} reduces to
\begin{equation}
    \frac{d\tilde z}{dt} = q (n-1)(\cot (q |\log\epsilon|)  + i\sigma)\frac{\tilde z}{\tilde r^2}
    \label{eq:dztildedt_1}
\end{equation}
Using the fact that $1-\chi = 1/n$ leads to
\begin{equation}
    \frac{d\tilde z}{dt} = q \frac{\chi}{1-\chi}(\cot (q |\log\epsilon|)  + i\sigma)\frac{\tilde z}{\tilde r^2}
    \label{eq:dztildedt_2}
\end{equation}

The first term in Eq.~\eqref{eq:dztildedt_2}, a real inverse mobility term, represents the repulsion (attraction) from (of) a spiral to a region of positive (negative) Gaussian curvature irrespective of the charge, which via the $\epsilon$ factor depends on its position. The second term in Eq.~\eqref{eq:dztildedt_2}, an imaginary inverse mobility term, represents orbital dynamics about the apex. Thus combining both terms, the mobility is complex for CGLE. In the limit $q\to0$, only the radial term survives, leading to
\begin{equation}
    \frac{d\tilde z}{dt} = \frac{\chi}{1-\chi}\frac{1}{|\log\epsilon|}\frac{\tilde z}{\tilde r^2}
\end{equation}
which represents the purely radial interaction as expected~\cite{vitelli2004anomalous}. As $|q\log\epsilon|$ increases from $0$ to $\pi/2$, the perpendicular term becomes increasingly important, causing the spiral motion to interpolate between movement along and perpendicular to the radial line, thereby enabling more complex spiral dynamics. This derivation provides the simplest analytic setting where geometry generates drift through the metric, and it anticipates the more general diffusion based mechanism developed below for arbitrary surfaces.

\subsection{Comparison with numerical simulations}
Performing simulations on a cone is challenging and requires care, for example matching the boundary conditions on either side of the wedge to ensure a smooth transition. Conveniently, there exists a well defined coordinate system called isothermal coordinates~\cite{gauss1822on,nelson1989statistical,vafa2022active} that avoids this technicality, as done in Refs.~\cite{vafa2023active}. Specifically, we define isothermal coordinates $z = re^{i \theta}$ by
\begin{align}
    \tilde z = \frac{z^{1-\chi}}{1-\chi} \label{eq:isothermal_cone}
\end{align}
Note that since the phase of $\tilde z$ ranges from $0$ to $2\pi(1-\chi)$, the phase of $z$ ranges from $0$ to $2\pi$.

In polar isothermal coordinates ($r, \theta$), the CGLE (Eq.~\eqref{eq:CGLE}) on a cone becomes
\begin{equation}
\frac{\partial\psi}{\partial t} = \psi + 4 r^{2\chi}(1 + i\alpha)\partial\bar\partial \psi  - (1 + i\beta)|\psi|^2 \psi
\end{equation}

In these coordinates, written in terms of $\alpha = -q$ (when $\beta = 0$), Eqs.~\eqref{eq:dztildedt_2} become
\begin{subequations}
\begin{align}
    \frac{dr}{dt} &= \alpha\chi\cot (\alpha |\log\epsilon|) \frac{1}{r^{1-2\chi}}\\
    \frac{d\theta}{dt} &= -\alpha\chi\sigma \frac{1}{r^{1-2\chi}}
\end{align}
\end{subequations}
Note that to leading order in $\alpha$ (corresponding to $\alpha \ll 1$), the above equations for $\sigma=-1$ reduce to
\begin{equation}
    |\log\epsilon| \, \frac{dr}{dt} + i \frac{d\theta}{dt} = \frac{1}{2}\partial_r[(1+i\alpha)r^{2\chi}]
    \label{eq:v_negative-charge}
\end{equation}
We will return to this equation in Sec.~\ref{sec:curved} when we consider general surfaces.

We simulate a single defect in isothermal coordinates to test our near field theory. An example of a defect with its trajectory is shown in Fig.~\ref{fig:cone_full}e as a reference for when the data is transformed to physical coordinates.

For the regime of moderate to high $\alpha = 0.5$, as used throughout in the paper, we observe distinct behaviors for positive ($\chi=0.25)$ and negative ($\chi = -0.25$) Gaussian curvature (Fig.~\ref{fig:cone_full}c,d). For positive Gaussian curvature, defects initially follow analytical predictions reasonably well over moderate distances. However, as they move further from the singularity, their behavior progressively diverges from the near field theory. While the simulated defect continues to spiral outward, the analytical trajectories settle into circular orbits. For negative Gaussian curvature, the behavior is markedly different. Rather than being attracted toward the singularity as the near field theory predicts, the simulated defect is instead slowly repelled away from the center. These deviations at larger parameter values reveal the limitations of the near field approximation when defects venture beyond the immediate vicinity of the singularity.

We further test our theory in the small parameter regime where the near field approximation is most valid ($\alpha = 0.1$, $\chi=\pm 0.1$). In this parameter regime, setting the fitting parameter $\epsilon = \tilde r/\tilde a = r/5.0$, the simulated trajectories and analytical predictions match remarkably well (Fig.~\ref{fig:cone_full}f,g). For both positive Gaussian curvature, where the defect is repelled, and negative Gaussian curvature, where the defect is attracted, the simulated defect trajectory closely follows the analytical solution. This strong agreement validates our theoretical framework in the appropriate limit where the assumptions of the near field theory hold.

\section{Gaussian bump induces stable orbits}

\begin{figure}[t!]
    \centering
    \includegraphics[width=0.95\linewidth]{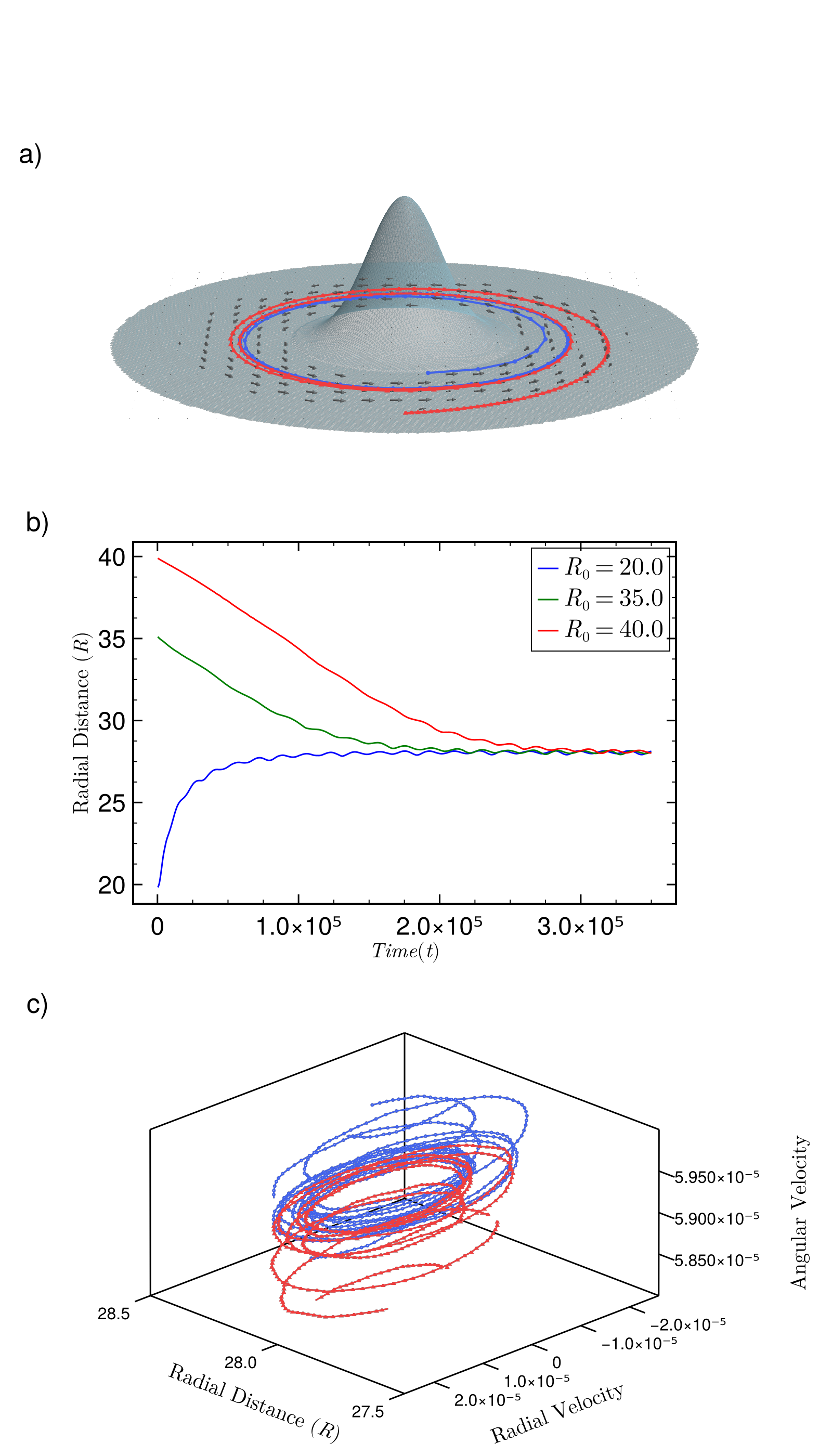}
    \caption{\textbf{Defect trajectories near a bump exhibit rotational motion due to nonlocal curvature coupling.} Two trajectories, near (blue) and further away (red) both converge and begin to orbit (a). In time, there is not an exact specific stable radius but small oscillators around $r^* \approx 9$ (b) in what appears to be a limit cycle (c).}
    \label{fig:nonlocal}
\end{figure}

Having validated our near field theory on cones, where the Gaussian curvature vanishes except for at the apex, we next investigate the effects of distributed curvature, which is not directly captured in our analytical treatment.

We simulate a defect initialized on a flat disk, with a localized curved bump positioned at a distance, where the bump is defined $h(x, y) = Ae^{-(x^2 + y^2)/(2\sigma^2)}$, with $A=25$ and $\sigma=6.25$. In order to ensure the surface is flat, we set $h=0$ when the height is less than the edge length in our mesh, this corresponds to a distance of $r = 19$ from the peak.

On a Gaussian bump, the total Gaussian curvature enclosed by any disk of finite radius remains positive, and for $\alpha = 0$, the defect experiences a purely repulsive radial force with no preferred radius.

When $\alpha\neq 0$ the defect settles into a stable orbit around the bump (Fig.~\ref{fig:nonlocal}a). Moreover, independent of initialization, defect trajectories eventually converge to the same stable orbital radius. Analysis of the radial evolution and phase space reveals that this is not a simple circular orbit, but rather a stable limit cycle characterized by small, persistent oscillations (Fig.~\ref{fig:nonlocal} b,c). Thus $\alpha$ is not merely stabilizing but is necessary for the orbit to exist.

The observation that these orbits act as attractors implies that geometric features can serve as localization traps. However, unlike negative Gaussian curvature, defects here are trapped in a dynamic steady state. The underlying mechanism for stable orbits can be explained as follows: the presence of Gaussian bump introduces a finite geometric length scale through its width $\sigma$, which is absent on a cone. This finite scale allows the $\alpha$ dependent spiral dynamics to balance curvature induced repulsion at a finite radius, producing a stable limit cycle. By contrast, the scale free geometry of a cone admits spiral motion but not orbital localization. The resulting orbital states therefore reflect the influence of geometry beyond the defect core, mediated by the extended spiral arms rather than by local Gaussian curvature alone. This behavior is a direct signature of a long range geometric force transmitted by the wave field.

\section{General curved geometry can be modeled by effective diffusion}
\label{sec:curved}
Our results thus far have been limited to specific curved surfaces, such as cones, bumps, and saddles. Each helped in building theory and understanding, while producing distinct defect behaviors such as directed motion, localization, and stable orbits. We now provide a unifying physical mechanism that explains all these observations in a common language.

We will show that in isothermal coordinates, the Laplace Beltrami operator takes a form in which the physical diffusion coefficient is multiplied by a spatially varying metric factor (the conformal factor). Although the physical diffusion coefficient remains uniform, in isothermal coordinates it appears as if the effective diffusion $D_{\text{eff}}$ varies in space. Thus curvature affects spiral dynamics indirectly: not by changing the physical diffusion coefficient, but by modifying the metric in a way that produces spatial variations in $D_{\text{eff}}$. Gradients of $D_{\text{eff}}$, together with the complex mobility set by the CGLE parameters, generate a geometric force that drives defect motion. This viewpoint elevates diffusion from a passive transport term to an active mediator through which geometry controls nonequilibrium dynamics.

\subsection{Curved geometry as varying diffusion coefficient}
For arbitrary surfaces, it is convenient to work in isothermal coordinates, where the intrinsic metric can be written as
\begin{equation}
    ds^2 = e^{\varphi(z, \bar z)} |dz|^2 \label{eq:ds2_isothermal}
\end{equation}
where $e^{\varphi(z, \bar z)}$ is known as the conformal factor.
For a cone,
\begin{equation}
     \varphi = -\chi\ln(z\bar z) . \label{eq:phi_cone}
\end{equation}
In wedge coordinates, note that, substituting Eqs.~\eqref{eq:isothermal_cone} and \eqref{eq:phi_cone} into Eq.~\eqref{eq:ds2_isothermal} leads to
\begin{equation}
     ds^2 = |d\tilde z|^2
\end{equation}
as expected.

In isothermal coordinates, the Gaussian curvature $K$ has a particularly simple form
\begin{equation}
    K\sqrt{g} = -2 \partial\bar\partial \varphi,
\end{equation}
where $\sqrt{g} = e^{\varphi}$ and partial derivatives $\partial\equiv (\partial_x - i\partial_y)/2$ and $\bar\partial\equiv (\partial_x + i\partial_y)/2$. In the case of a cone,
\begin{equation}
    K\sqrt{g} = 2\pi\chi \delta^{(2)}(z)
\end{equation}
As expected, the curvature is localized at the apex, and upon integration, leads to a deficit angle of $2\pi\chi$.

In isothermal coordinates, the CGLE becomes
\begin{align}
\frac{\partial \psi}{\partial t} &= \psi + 4 e^{-\varphi}(1 + i\alpha)\partial\bar\partial \psi - (1 + i\beta)|\psi|^2 \psi \nonumber\\
&= \psi + D_{\text{eff}}(1 + i\alpha)(\partial_x^2 + \partial_y^2) \psi - (1 + i\beta)|\psi|^2 \psi
\end{align}
where $D_{\text{eff}} = e^{-\varphi}$. Thus it is as if we have position dependent diffusion coefficient $D_{\text{eff}}$ on a flat plane. Notice that in the case of a cone, the leading order $\alpha$ velocity components (Eq.~\eqref{eq:v_negative-charge}) becomes
\begin{align}
    2\left(|\log\epsilon| \frac{dr}{dt} + i \frac{d\theta}{dt}\right) &= \partial_r[(1+i\alpha)r^{2\chi}] \nonumber \\
    &= \partial_r [(1 + i\alpha)D_{\text{eff}}],
\end{align}
i.e.,
\begin{subequations}
\begin{align}
    2|\log\epsilon| \frac{dr}{dt} &= \partial_r D_{\text{eff}} \\
    2 \frac{d\theta}{dt} &= \alpha\partial_r D_{\text{eff}},
\end{align}
\end{subequations}

The simple local form above should be regarded as an approximation conjectured to be valid for slowly varying $D_{\text{eff}}$ profiles. For more general profiles of the effective diffusion, we expect the defect velocity to take the general form
\begin{equation}
    \frac{d\mathbf{x}}{dt} = A(\mathbf{x})\,\nabla D_{\text{eff}}(\mathbf{x}) + B(\mathbf{x})\,\nabla^{\perp} D_{\text{eff}}(\mathbf{x}).
\end{equation}
Equivalently, in isothermal coordinates, one may write
\begin{equation}
    v^z(z) = \Big(a(z) + i\,b(z)\Big)\,e^{-\varphi(z)}\,\bar\partial\varphi(z).
\end{equation}
where $e^{-\varphi} = g^{z\bar z}$ is the inverse metric component in isothermal coordinates. Here, $a(z)$ and $b(z)$ are inverse mobility coefficients which we expect to depend on the geometric potential $\varphi$, curvature $K$, and the model parameters $\alpha$ and $\beta$, i.e., we conjecture that
\begin{align}
    a(z) &= \tfrac{q}{2}\;C_r[\varphi(z), K(z); \alpha, \beta],\\
    b(z) &= \tfrac{q}{2}\;C_t[\varphi(z), K(z); \alpha, \beta] .
\end{align}
Writing the coefficients in terms of $\varphi$ and $K$ makes the metric dependence explicit since $D_{\mathrm{eff}}=e^{-\varphi}$ and $K$ is fixed by the metric. In the near field limit on a cone, these functionals recover $C_r\to\cot(q|\log\epsilon(r)|)$ and $C_t\to\sigma$.

We test the $D_{\text{eff}}$ based ansatz in the next subsection using numerical experiments with imposed diffusion profiles.

\subsection{An intuitive picture of diffusion gradient}

\begin{figure}
    \centering
    \includegraphics[width=1\linewidth]{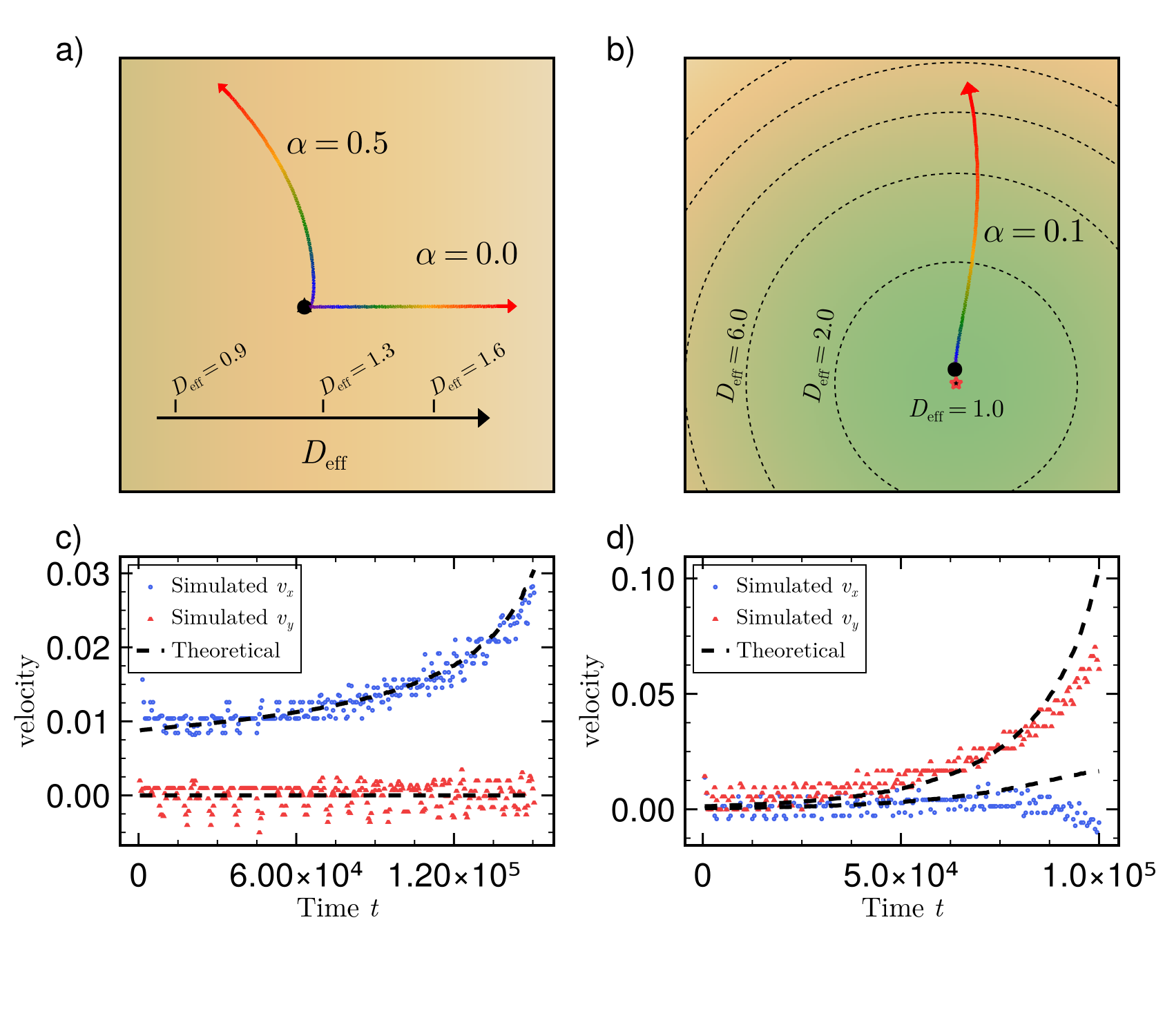}
    \caption{\textbf{Analytical predictions match simulations for simple $D_{\text{eff}}$ gradients.} (a) Trajectory of spiral defects in a linear $D_{\text{eff}}$ gradient. (b) Trajectory in a quadratic $D_{\text{eff}}$ gradient. (c) Comparison of simulated and analytical (dashed lines) velocity components for the linear gradient case, showing excellent agreement between theory and simulation when $\alpha=0$. (d) Velocity comparison for the quadratic gradient case match for $\alpha=0.1$.}
    \label{fig:lin_quad}
\end{figure}

This viewpoint suggests that curvature is just one mechanism for generating such gradients, and that any system with generic spatial variations in $D_{\text{eff}}$ regardless of their origin should exhibit analogous defect dynamics. To test this hypothesis, we remove curvature entirely and impose diffusion gradients directly on a perfectly flat surface. If defects behave similarly in this simplified flat system, it would confirm that gradients in effective diffusion are sufficient to drive the observed phenomena, and that their specific geometric origin is secondary.

The simplest test case is a flat plane with a linear gradient in $D_{\text{eff}}$ (Fig.~\ref{fig:lin_quad}a). When $\alpha = 0$, the defect moves in a straight line from regions of low to high diffusion, consistent with non spiraling defects on curved surfaces~\cite{vafa2022defectAbsorption,vafa2023active}. When $\alpha \neq 0$, an additional transverse velocity emerges perpendicular to the gradient, with direction set by spiral chirality and magnitude determined by gradient strength and $\alpha$, not only for linear gradient (Fig.~\ref{fig:lin_quad}a) but also a quadratic with $D_{\text{eff}}  = a(x^2 + y^2) +1.0$ (Fig.~\ref{fig:lin_quad}b). Comparing analytical predictions (dashed) with simulations shows agreement for linear gradients with $\alpha=0.0$ (Fig.~\ref{fig:lin_quad}c) and quadratic diffusion (Fig.~\ref{fig:lin_quad}d).

\begin{figure*}
    \centering
    \includegraphics[width=\linewidth]{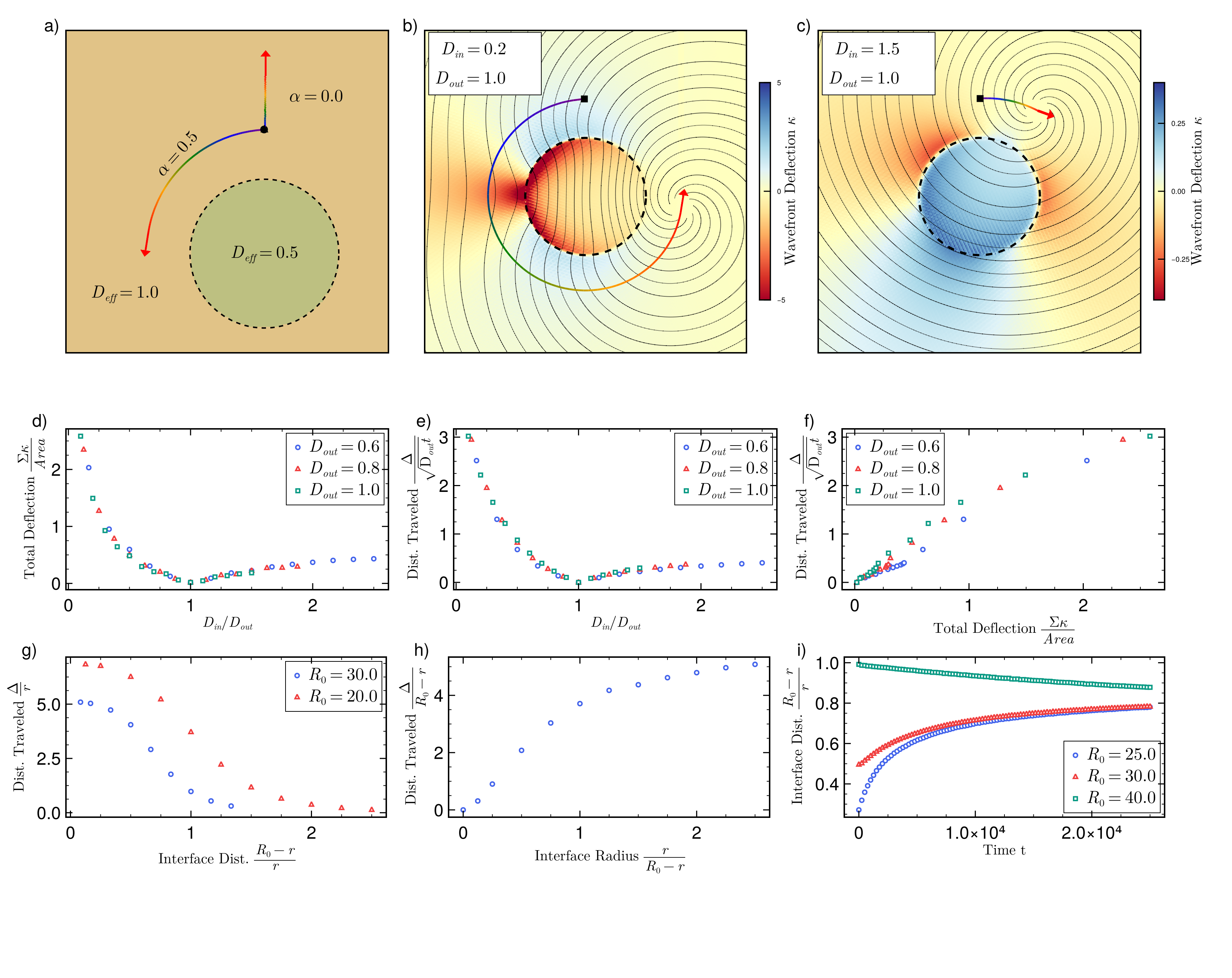}
    \caption{\textbf{Spatial gradients in the diffusion coefficient drive defect motion on flat surfaces.} (a) Defect trajectories (lines) under a circular step diffusion profile. Panels (b) and (c) show the deflection of the spiral wave front (see Eq.~\eqref{eq:bending}): when the inner diffusion coefficient is less than the outer value, the wavefront bend inwards which propels defects anticlockwise (b); on the other hand, reversing the $D_{in}$ and $D_{out}$ ratio causes opposite behavior (c). In (b) and (c), $\kappa$ is normalized by the deflection for a spiral wave located at the same position but with no change in $D_\text{eff}$. The total deflection (d), and the distance traveled (e) follow the same curve against the ratio of $D_{\text{eff}}$ and show a proportionality (f). Initializing the defect further from the reduced D region causes it to travel less (g) as does reducing the size of the region (h), but it eventually comes and settles into the same orbital radius (i).}
    \label{fig:vardiff}
\end{figure*}

A more controlled version of the same setup is presented in Fig.~\ref{fig:vardiff}. Here, instead of a linear gradient, a localized circular region is set with $D_{\text{eff}}= D_{\text{in}}$ while the surrounding region is fixed at $D_{\text{eff}}=D_{\text{out}}$. This is a similar yet simplified situation as the Gaussian bump geometry. This configuration allows us to gradually increase the diffusion ratio ($D_{\text{in}}/D_{\text{out}}$), thereby increasing the gradient at the interface and thus the defect propulsion.

The origin of this propulsion lies in wavefront deflection. When a spiral wave encounters a region with lower effective diffusion, the wavelength, as measured in isothermal coordinates, decreases in that region, with the opposite behavior for higher $D_{\text{eff}}$. Because the wavefront extends radially outward from the defect core, a diffusion gradient across the wavefront causes one side to compress or stretch relative to the other. This asymmetry bends the wavefront. We define the wavefront as contours of constant phase and quantify its deflection by measuring the curvature of these contour lines. The defect then moves to relieve this induced curvature, with the direction determined by the sign of the defect and the magnitude of deflection.

Fig.~\ref{fig:vardiff}b,c illustrate this effect for two diffusion ratios. When $D_{\text{in}}/D_{\text{out}} = 0.2$, the wavelength reduces sharply inside the circular region, producing strong wavefront deflection that drives anticlockwise motion. The region inside has large negative deflection, which is when the wavefront bends in a concave shape. In contrast, $D_{\text{in}}/D_{\text{out}} = 1.5$ shows positive deflection inside the region, which is the result of the wavelength expanding and bending the wavefront in the opposite direction. This produces clockwise motion. The color map shows the normalized deflection, which is the local wavefront curvature divided by that of a defect at the same position in a system without a change in $D_{\text{eff}}$. With phase $\phi$, the wavefront curvature is defined:
\begin{equation}
    \kappa = \frac{\partial_{xx}\phi \, \partial_y\phi^2 - 2\partial_{xy}\phi \, \partial_x\phi \, \partial_y\phi + \partial_{yy}\phi \, \partial_x\phi^2}{(\partial_x\phi^2 + \partial_y\phi^2)^{3/2}}
    \label{eq:bending}
\end{equation}

We quantify this relationship by plotting the total deflection, defined as $1/A \int \kappa dA$, and the distance traveled by the defect across a range of diffusion ratios, each for three separate values of $D_{\text{out}}$. All curves follow the same trend and produce curves that overlap almost perfectly, indicating that the absolute values of diffusion matter less than their ratio (Fig.~\ref{fig:vardiff}d,e). In both plots, the curves reach a minimum when the ratio is 1 and then increase as the ratio deviates in either direction. For both, there is a strong asymmetry for reduced versus enhanced $D_{\text{eff}}$ in the circular region. Most notable is that the deflection and displacement plots are nearly identical in shape. When plotted against each other, a linear dependence emerges where the defect displacement is directly proportional to accumulated wavefront bending (Fig.~\ref{fig:vardiff}f).

Two additional parameters influence the magnitude of defect motion, the initial distance from the interface and the size of the circular region (Fig.~\ref{fig:vardiff}g). Defects initialized closer to the interface experience stronger deflection and travel farther. Similarly, increasing the radius of the circular region also leads the defects to travel further, but only up to a threshold. After a certain radius, the gain in defect propulsion reduces (Fig.~\ref{fig:vardiff}h). Despite these variations, plotting $r(t)$ shows that all trajectories converge to the same stable orbital radius regardless of initial distance (Fig.~\ref{fig:vardiff}i). This mirrors the behavior observed near the Gaussian bump, where defects initialized at different positions also settled into orbits.

From these results, we can identify three distinct forces that govern defect motion. First, the $\alpha = 0$ control case reveals a nonchiral force that drives defects toward or away from regions of changing diffusion. The emergence of tangential motion when $\alpha > 0$, along with the reversal of orbital direction between opposite ratio of $D_{\text{eff}}$, indicates a second force, which is the tangential chiral component proportional to wavefront deflection. The linear relationship between total deflection and distance traveled confirms that this force scales directly with wavefront bending. Finally, the tendency for defects to move towards a stable orbit, from both directions, points to a third force. This is also a radial component, but one that arises from the rotational motion itself. Together, these three forces combine to determine the overall defect trajectory.

The correspondence between surface geometry and spatially varying diffusion provides a unified framework for understanding defect dynamics on curved surfaces. Curved regions modify the local effective diffusion, generating gradients that drive defect motion either toward or away from these regions depending solely on the gradient direction. This mechanism accounts for the attraction observed for positive Gaussian curvature and the repulsion for negative Gaussian curvature, which correspond to regions of reduced and enhanced effective diffusion at the center, respectively. The chirality reversal between positive and negative curvature emerges naturally from how spiral wavefront bend when encountering these diffusion gradients. On a general surface, an exact form of $D_{\text{eff}}$ is not readily obtained; however, the force based intuition from this section is still expected to hold. In the next section we turn our attention to precisely these sorts of surfaces and find that there too, curvature plays a noticeable long range role, even in the presence of strong defect pair interactions.

\section{Curvature impact on multi defect interactions \label{results2}}

\begin{figure*}
    \centering
    \includegraphics[width=0.95\linewidth]{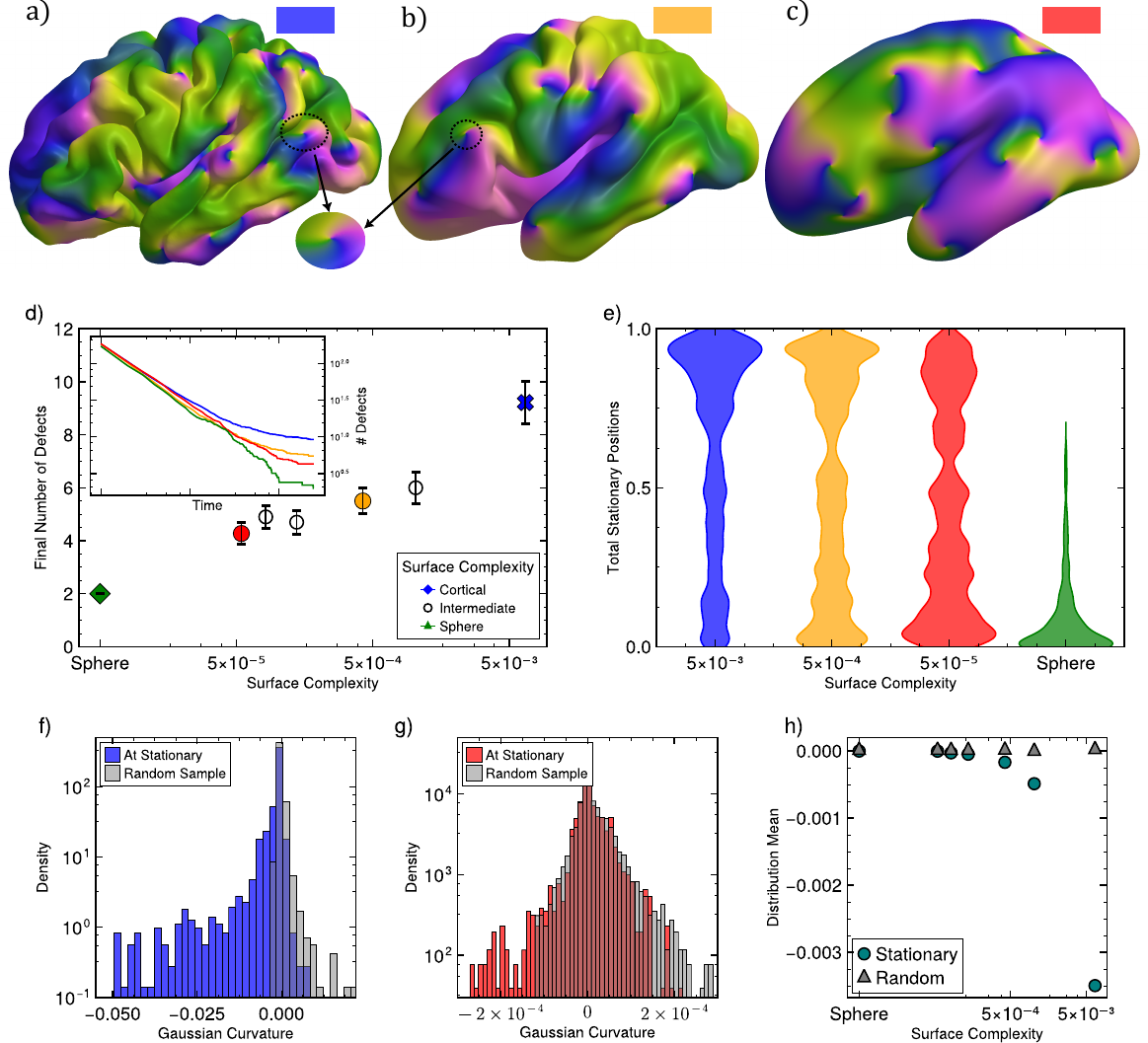}
    \caption{\textbf{Surface smoothing dilutes curvature effects.} Visualization of the cortical surfaces without smoothing (a), with gradually increased smoothing (b,c). Final defect number as a function of surface complexity, where more folded surfaces retain more defects (d). Inset: defect number versus time for each geometry. The sphere (green diamond) follows a power law decay to two defects, while the cortical surface (blue cross) plateaus at $\approx 10$. Distribution for the fraction of a defect lifetime it spends stationary (e) and histograms of Gaussian curvature at locations defects were stationary in our simulations on the surfaces in (a) and (c) (colored) compared with a random selection of points (gray) (f, g). The shift towards negative Gaussian curvature is captured by the mean of the distribution (h). The color scheme in all plots follows the box presented above each representative snapshot.}
    \label{fig:brain}
\end{figure*}

The mechanism we have identified acts at long range through effective diffusion gradients, capable of driving defects across entire surfaces toward or away from curved regions, similar to experimental observations on the brain~\cite{ye_brain-wide_2023}. At short range, however, pairwise interactions dominate. When two defects of opposite (same) charge approach each other, they attract and annihilate (repel). Curvature can modify the strength of these interactions, but it does not alter the outcome. Whether the long range geometric forces that we have identified can compete with short range pairwise effects in dense systems remains unclear.

We investigate this by initializing defects at high density on the surface of a human brain. This geometry is full of folds, ridges, and valleys that produce strong and spatially varying curvature. We compare the resulting dynamics to those on a sphere of equivalent surface area, where the isotropy of the surface prevents geometric trapping. To isolate the role of geometric surface complexity, we also analyze a series of intermediate geometries where the folds are gradually smoothed away, interpolating between the cortical surface and the sphere (Fig.~\ref{fig:brain}, panels a-c). In order to have a measurement of complexity, we use the standard deviation of the Gaussian curvature distribution. Importantly, the parameter pair $\alpha = 0.5$, $\beta = 0.0$ produces frozen states where defects become immobile before annihilating; to ensure defects remain mobile and can annihilate given sufficient time, we use $\alpha = 0.5$, $\beta = 0.25$ throughout this section. We first confirm on the sphere that this parameter choice produces the expected behavior where defects interact through pairwise forces and progressively annihilate until only two remain.

Plotting the final defect number against geometric complexity reveals a clear trend: as the surface is smoothed to a sphere the number of defects gradually reduce Fig.~\ref{fig:brain}d. The sphere reaches the minimum of two, while the realistic cortical surface retains more defects even after long times. The intermediate geometries fall between these extremes, with defect survival increasing with the surface complexity. The time evolution of defect number (inset Fig.~\ref{fig:brain}e) shows how this difference emerges. On the sphere, the decay follows a power law for almost the entire time span, indicating that there is no curvature effect. While time series for the realistic brain and intermediate surfaces begin similarly, they quickly deviate from this power law. Here, when density is high, the defects are strongly affected by pairwise interactions and rapidly annihilate. As the density is reduced, the decline slows down as encounters become rare. This systematic dependence on geometric complexity indicates that curvature is responsible for the enhanced defect retention, but it does not reveal whether defects on these surfaces are merely moving more slowly, leading to delayed annihilation, or becoming trapped altogether.

To distinguish between slow motion and trapping, we plot the fraction of time each defect spends stationary across four geometries Fig.~\ref{fig:brain}f, defining stationary as times when a defect has velocity below the detection threshold for our simulation scheme for multiple successive recorded states. The distributions are completely different. On the sphere, nearly all defects remain motile for the majority of the time with the distributions sharply peaked near zero, indicating that defects annihilate through continuous pairwise encounters. The cortical surface shows the opposite distribution, with many defects spending $90\%$ or more of their time immobile. Two intermediate geometries show this transition, gradually shifting toward motile defects. As the folds are smoothed away, the peak at high stationary fraction diminishes and the distribution shifts toward lower values. At the smoothest intermediate level (Fig.~\ref{fig:brain}(c)), the distribution is already trending towards that of the sphere. These defects are not slowly drifting toward annihilation but are trapped in place for extended periods of time. The more folded the surface, the more effectively it immobilizes defects and prevents annihilation. In physical terms, smoothing reduces the gradients of the effective diffusion and thereby weakens the geometric force that otherwise localizes defects.

Trapped defects preferentially localize in regions of negative Gaussian curvature. We quantify this by computing histograms of Gaussian curvature at stationary positions and comparing these to random samples drawn from the entire surface. On the cortical surface (Fig.~\ref{fig:brain}(g)), the distribution for stationary defects is shifted relative to the random sample (Fig.~\ref{fig:brain}(f)). The peaks roughly align, but the negative tail is more pronounced. The smoothest intermediate geometry (Fig.~\ref{fig:brain}(c)) shows a much weaker version of this effect (Fig.~\ref{fig:brain}(g)), with the distributions nearly overlap. While the geometry did still retain a few defects at the end of the simulation, there is almost no difference between the random selection at the stationary locations. Plotting the mean of these distributions against surface complexity (Fig.~\ref{fig:brain}(h)) quantifies this trend for Gaussian curvature. As the surface becomes more folded, the average curvature at stationary positions shifts progressively toward negative values. This preferential localization, though subtle in magnitude, is significant over long times. It demonstrates that curvature induced trapping provides a stabilization mechanism in multi defect systems, allowing geometry to influence defect dynamics even when pairwise forces would otherwise dominate.

These findings have important implications for systems with ongoing defect creation and annihilation, as is more realistic in biological tissues. In such systems, the balance between creation and annihilation rates typically sets a steady state defect organization where defects do not preferentially cluster. Our results suggest that on curved surfaces, trapping at regions of negative Gaussian curvature would shift this balance, increasing steady state densities in these regions compared to others. Such behavior is in line with our results with single defects, where it was seen that bound and orbiting states could be formed by some geometric conditions.

\section{Discussion}
Our results reveal a direct and universal mechanism by which surface geometry governs the dynamics of spiral waves and their topological defects. The key insight is that curvature does not simply distort spatial patterns but acts through the Laplace Beltrami operator as spatially varying effective diffusion. This coupling between curvature, diffusion, and phase transforms geometry from a passive boundary condition into an active field that generates forces on defects. In this sense, the work identifies a new class of nonequilibrium forces: geometric forces mediated by diffusion and transmitted by the wave field. The resulting dynamics, including ballistic motion, orbital trajectories, and long lived trapped states, arise from purely geometric properties and therefore transcend the details of any specific medium.

The same principle explains why curved biological surfaces often display stable and localized spiral activity. In the brain, the folded geometry of the cortex creates curvature gradients at multiple spatial scales. Our analysis shows that such gradients can stabilize spiral waves, prolong their lifetimes, and bias their spatial distribution toward regions of negative Gaussian curvature. The implication is profound: cortical folding is not only a mechanical or developmental feature but a physical parameter that shapes the repertoire of available brain dynamics. In this view, large scale brain activity patterns are not determined solely by synaptic connectivity or local excitability but also by the underlying geometry that constrains how waves can propagate and interact.

This geometric perspective provides a unifying framework for linking structure and function in the brain. It complements network based descriptions by introducing a continuous, physics based route through which anatomical form can modulate dynamical state. The framework also offers a mechanistic explanation for recent observations that the spectrum of cortical dynamics follows geometric eigenmodes~\cite{pang2023geometric} and that spiral waves preferentially align with the folding pattern of the cortex~\cite{ye_brain-wide_2023, xu_spatiotemporal_2025}. In both cases, the underlying cause may be the same: the modulation of effective diffusion by curvature and topology.

Beyond the cortex, the mechanism we identify applies broadly to oscillatory, chemical, and active matter systems evolving on curved substrates. In all such systems, geometry can generate nonlocal coupling and emergent forces that reshape pattern dynamics. This opens new opportunities to design curvature controlled patterning in synthetic materials, active films, and catalytic surfaces. It also suggests new experimental strategies: by systematically altering surface curvature, one could steer the formation, motion, or annihilation of defects in real time, creating a new class of geometrically programmable dynamical media.

Looking forward, several directions of interest emerge. First, the geometric forces identified here can be combined with heterogeneous or anisotropic coupling to explore how geometry and connectivity interact to shape complex brain dynamics. Second, extending this framework to incorporate the effect of mean curvature is an exciting direction given the prominence of intricate folding of the cortical surface. Third, incorporating growth and remodeling can reveal feedback loops in which activity patterns influence, and are in turn constrained by, evolving geometry, an avenue that may shed light on the co development of brain form and function. Finally, the correspondence between curvature and effective diffusion suggests the possibility of defining geometric order parameters for non equilibrium systems, enabling a systematic classification of curvature driven pattern formation.

Taken together, our findings identify curvature as a universal organizing principle for non equilibrium dynamics on complex surfaces. In the brain, this coupling transforms cortical geometry from a static scaffold into a physical field that actively shapes neural activity. Recognizing geometry as a fundamental dynamical variable opens a new direction for the physics of living systems and establishes a direct physical link between the shape of the brain and the patterns of thought and perception it supports.

\begin{acknowledgments}
We thank Mehran Kardar for helpful discussions. 
F.V. acknowledges funding from the Novo Nordisk Foundation (grant No.~NNF18SA0035142). 
M.C.P. thanks Villum Fonden (Grant No.~69081) for their support.
A.D. acknowledges funding from the Novo Nordisk Foundation (grant No.~NNF18SA0035142 and NERD grant No.~NNF21OC0068687), Villum Fonden (Grant No.~29476), and the European Union (ERC, PhysCoMeT, 101041418).
The Tycho supercomputer at the University of Copenhagen supported this work.
\end{acknowledgments}

\bibliography{references}

\end{document}